\documentclass[10pt,aps,prl,twocolumn,preprintnumbers,amsmath,amssymb,floatfix,showpacs,citeautoscript]{revtex4-1}
\usepackage[latin1]{inputenc}
\usepackage[T1]{fontenc}
\usepackage[english]{babel}
\usepackage[pdftex]{graphicx}
\usepackage{amsmath}
\usepackage{times}
\usepackage[usenames,dvipsnames,svgnames,table]{xcolor}
\usepackage[colorlinks,plainpages=false,linkcolor=blue,urlcolor=blue,citecolor=blue,pdfpagemode=UseNone]{hyperref}

\renewcommand{\AA}{\text{\r{A}}}

\newcommand\Multp{\cdot}
\newcommand\Vek[1]{\vec{#1}}

\begin{document}

\title
{
\boldmath
Fundamental difference in the electronic reconstruction of infinite-layer vs.\ perovskite neodymium nickelate films on SrTiO$_3$(001)
}

\author{Benjamin Geisler}
\email{benjamin.geisler@uni-due.de}
\affiliation{Department of Physics and Center for Nanointegration (CENIDE), Universit\"at Duisburg-Essen, Lotharstr.~1, 47057 Duisburg, Germany}
\author{Rossitza Pentcheva}
\email{rossitza.pentcheva@uni-due.de}
\affiliation{Department of Physics and Center for Nanointegration (CENIDE), Universit\"at Duisburg-Essen, Lotharstr.~1, 47057 Duisburg, Germany}

\date{\today}

\begin{abstract}
Motivated by recent reports of superconductivity in Sr-doped NdNiO$_2$ films on SrTiO$_3$(001) [Nature (London) \textbf{572}, 624 (2019)],
we explore the role of the polar interface on the structural and electronic properties of
NdNiO$_n$/SrTiO$_3$(001) ($n=2,3$) by performing
first-principles calculations including a Coulomb repulsion term. For infinite-layer
nickelate films ($n=2$), electronic reconstruction drives the surprising emergence of
a two-dimensional electron gas (2DEG) at the interface involving a strong occupation of the Ti $3d$ states.
This effect is more pronounced than in LaAlO$_3$/SrTiO$_3$(001)
and accompanied by a substantial reconstruction of the Fermi surface:
a depletion of the self-doping Nd $5d$ states
and an enhanced Ni $e_g$ orbital polarization reaching up to $35\%$ at the surface,
reflecting a single hole in the $3d_{x^2-y^2}$ states, i.e.,
cuprate-like behavior.
In contrast, no 2DEG forms for perovskite films ($n=3$)
or if a single perovskite layer persists at the interface.
We show that the topotactic reaction from the perovskite to the
infinite-layer phase is confined to the
nickelate film, whereas the SrTiO$_3$ substrate remains intact.
\end{abstract}

\maketitle

% \section{Introduction}

Interface polarity plays a key role in transition metal oxide heterostructures,
since it can drive the emergence of physical properties that are absent in the respective bulk compounds~\cite{Mannhart:10, Hwang:12, Ohtomo:02}.
A prominent example is the LaAlO$_3$/SrTiO$_3$(001) system (LAO/STO~\cite{Ohtomo:2004, Nakagawa:06}),
in which a correlated two-dimensional electron gas (2DEG)
is formed at the interface by occupation of Ti $3d$ conduction band states
beyond 4 unit cells of LAO~\cite{Thiel:06, PentchevaPickett:09}
that shows intriguing physics, for instance, superconductivity~\cite{Reyren:07}.

The very recent observation of superconductivity in Sr-doped NdNiO$_2$ films grown on STO(001)
by Li \textit{et al.}~\cite{Li-Supercond-Inf-NNO-STO:19}
has sparked considerable interest in infinite-layer ($AB$O$_2$) nickelates.
Most of the theoretical efforts to explain the phenomenon so far have concentrated on the electronic properties of bulk compounds~\cite{Nomura-Inf-NNO:19, JiangZhong-InfNickelates:19, Sakakibara:19, JiangBerciuSawatzky:19, Botana-Inf-Nickelates:19, Choi-Lee-Pickett-4fNNO:20},
particularly on the role of the Nd $5d$ states~\cite{Sawatzky-NNO:19, NNO-SelfDopingDesign-d9-Arita:20, NNO-SC-Thomale:20}.
Their self-doping impedes the emergence of a formal Ni$^{1+}$ ($3d^9$) valence state
that would render the nickelate close to cuprates.
Notably, superconductivity could not be confirmed experimentally in Sr-doped bulk NdNiO$_2$~\cite{Li-NoSCinBulkDopedNNO:19}.
This raises a question about the role of the interface to the substrate that has hardly been addressed so far.

Here we explore the impact of the polar interface
% to the STO(001) substrate
on the structural and electronic properties
of NdNiO$_n$/STO(001) ($n=2,3$) by performing
first-principles calculations including a Coulomb repulsion term
and find it to play a decisive role.
While polar discontinuities exist at the interface and the surface in both the infinite-layer [$n=2$; Nd$^{3+}/$(TiO$_2$)$^{0}$] and the perovskite [$n=3$; (NdO)$^{1+}/$(TiO$_2$)$^{0}$] films,
we find the accommodation mechanism to be highly different.
The simulations unravel a clear 2DEG formation for $n=2$ due to a strong occupation of interfacial Ti, predominantly $3d_{xy}$ states, more pronounced than in LAO/STO(001).
The 2DEG forms already for a single unit cell of infinite-layer nickelate on STO(001).
The electronic reconstruction is accompanied by a substantial reconstruction of the Fermi surface
as well as strong ionic relaxations and Ni valence modulations.
Particularly, we find the Nd $5d$ states to be pushed above the Fermi energy.
In conjunction with a high Ni~$e_g$ orbital polarization reaching up to $35\%$ at the surface,
i.e., a single hole in the $3d_{x^2-y^2}$ states,
this renders the film geometry close to $3d^9$ cuprates, distinct from bulk.
In sharp contrast, for $n=3$ the Ti $3d$ states remain empty,
and the polarity mismatch is exclusively accommodated
by electrostatic doping of the nickelate, accompanied by ionic relaxations.
Finally, we analyze the oxygen deintercalation reaction and show why the reduction is confined to the nickelate film.
Interestingly, this reveals the possible persistence of a single perovskite layer at the interface,
which simultaneously suppresses the 2DEG formation and the Nd $5d$ self-doping.
These insights may be of key importance for a fundamental understanding of superconductivity in infinite-layer nickelates
and in polar materials on nonpolar substrates in general.

\textit{Methodology. ---}
We performed first-principles calculations in the framework
of density functional theory~\cite{KoSh65} (DFT)
as implemented in the Quantum ESPRESSO code~\cite{PWSCF}.
The generalized gradient approximation was used for the exchange-correlation functional  
as parametrized by Perdew, Burke, and Ernzerhof~\cite{PeBu96}.
Static correlation effects were considered within the DFT$+U$ formalism~\cite{QE-LDA-U:05}
employing $U=4$~eV on Ni and Ti sites,
in line with previous work~\cite{Liu-NNO:13, Botana-Inf-Nickelates:19, Geisler-LNOSTO:17, WrobelGeisler:18, GeislerPentcheva-LNOLAO:18, GeislerPentcheva-LNOLAO-Resonances:19}.
\begin{figure*}
	\centering
	\includegraphics[]{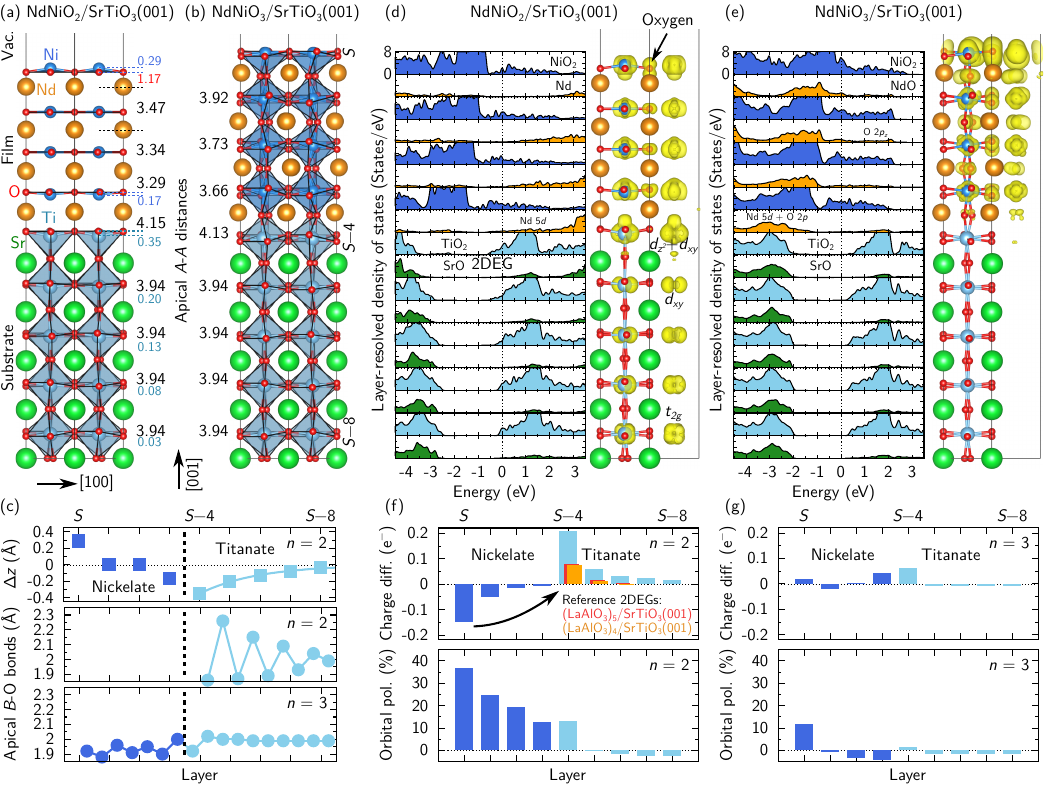}
	\caption{(a, b)~Optimized geometry of (NdNiO$_n$)$_4$/STO(001) ($n=2$: infinite-layer; $n=3$: perovskite). (c)~The vertical displacements $\Delta z$ of the $B$ site ions from their respective oxygen layers are substantial for $n=2$; for $n=3$, they are negligible (not shown). The apical $B$-O bond lengths reflect the highly distinct ionic relaxation in the two systems. (d, e)~Corresponding layer-resolved density of states and distribution of electron density (integrated between $-0.7$~eV and $E_\text{F}$), visualizing the fundamental differences in electronic reconstruction. The strong oxygen contributions to the electron density in the nickelate region (absent in STO) highlight its covalent nature. (f, g)~Layer-resolved charge difference at Ni and Ti sites relative to the respective bulk compounds and Ni and Ti $e_g$ orbital polarization of (NdNiO$_n$)$_4$/STO(001). The former show the valence modulations arising particularly for $n=2$. A positive orbital polarization indicates a preferential $3d_{z^2}$ occupation. For reference, the Ti charge differences are provided for the paradigmatic (LAO)$_4$/STO(001) and (LAO)$_5$/STO(001) 2DEG systems.\vspace*{-1.0em}}
	\label{fig:Structures-ElStr}
\end{figure*}
We model (NdNiO$_n$)$_m$/STO(001) ($n=2,3$) by using $\sqrt{2}a \times \sqrt{2}a$ supercells with two transition metal sites per layer to account for octahedral rotations,
strained to the STO substrate lattice parameter $a=3.905~\AA$.
The supercells contain $5$ layers of STO substrate and $m=1$-$6$ layers of nickelate film,
subsequently doubled to obtain symmetric slabs (the figures only show half of the supercell).
The vacuum region spans $20~\AA$.
We focus here on structural and electrostatic effects and therefore discuss nonmagnetic results in the following~\cite{Sawatzky-NNO:19, Li-Supercond-Inf-NNO-STO:19},
noting that spin-polarized calculations with ferromagnetic or A- and G-type antiferromagnetic order in the nickelate films resulted in qualitatively identical results with respect to 2DEG formation for both $n=2$ and $n=3$.
Wave functions and density were expanded into plane waves up to cutoff energies of $45$ and $350$~Ry, respectively.
Ultrasoft pseudopotentials~\cite{Vanderbilt:1990}
as successfully employed in previous work~\cite{GeislerPopescu:14, Geisler-Heusler:15, GeislerFePcHSi:19, Geisler-LNOSTO:17, WrobelGeisler:18, GeislerPentcheva-LNOLAO:18, GeislerPentcheva-LNOLAO-Resonances:19},
were used in conjunction with projector augmented wave datasets~\cite{PAW:94}.
The Nd $4f$ electrons are frozen in the core~\cite{Liu-NNO:13, Nomura-Inf-NNO:19};
their explicit treatment leads to qualitatively similar results.
We used a $12 \times 12 \times 1$ Monkhorst-Pack $\Vek{k}$-point grid~\cite{MoPa76}
and $5$~mRy Methfessel-Paxton smearing~\cite{MePa89} to sample the Brillouin zone.
The ionic positions were accurately optimized, reducing ionic forces below $1$~mRy$/$a.u.

\textit{Ionic relaxation. ---}
We focus our discussion on (NdNiO$_n$)$_4$/STO(001) heterostructures with a representative film thickness of 4 unit cells.
%, as displayed in Figs.~\ref{fig:Structures-ElStr}(a) and~(b).
%
The lattice parameters of bulk NdNiO$_2$ ($a = 3.92$, $c = 3.28~\AA$~\cite{Hayward:03, Nomura-Inf-NNO:19})
imply that the film is subject to compressive strain if grown epitaxially on STO(001) [$a = 3.905~\AA$, Fig.~\ref{fig:Structures-ElStr}(a)].
Consequently, the infinite-layer film expands vertically ($\sim 2.6\%$).
Interestingly, we find that this expansion is not homogeneous; instead, the distances between the distinct Nd layers increase continuously from the interface ($3.29~\AA$) to the surface ($3.47~\AA$).
For reference, the value for strained bulk NdNiO$_2$ is $3.29~\AA$.
An analogous trend is observed for the perovskite film [$n=3$, tensile strain, Fig.~\ref{fig:Structures-ElStr}(b)] with even larger distances.
The Nd-Sr distance at the interface is particularly enhanced for the infinite-layer film ($4.15~\AA$), slightly more pronounced than for the perovskite film ($4.13~\AA$).
We associate this result with the electrostatic doping,
similar to the previously observed enhanced La-Sr distance across
the $n$-type LaNiO$_3$/STO(001) interface ($\sim\!4.06~\AA$)~\cite{Geisler-LNOSTO:17, ZhangKeimer:14, Hwang:13}.

\begin{figure*}
	\centering
	\includegraphics[]{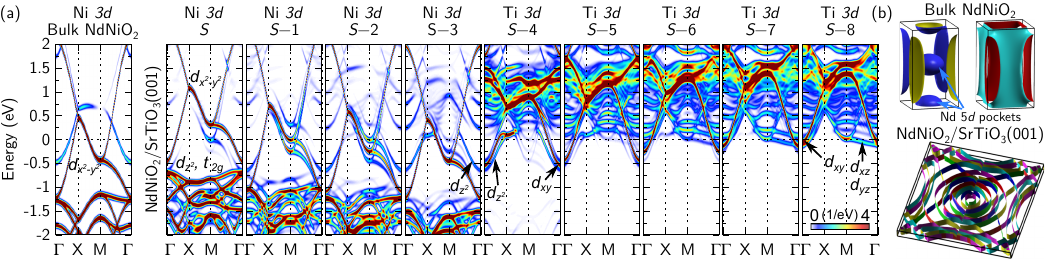}
	\caption{(a)~Band structures ($\Vek{k}$-resolved densities of states, projected on Ni $3d$ and Ti $3d$ orbitals in different layers) highlight the interfacial 2DEG formation in (NdNiO$_2$)$_4$/STO(001). For reference, a panel for bulk NdNiO$_2$ is provided. The orbital characters are denoted. (b)~The film Fermi surface is composed predominantly of Ni $3d_{x^2-y^2}$ and $\Gamma$-centered Ti $3d$ (2DEG) sheets and thus considerably reconstructed with respect to the Fermi surface of bulk NdNiO$_2$, which is shown here for a comparable $\sqrt{2} \times \sqrt{2}$ cell and closely resembles that of LaNiO$_2$~\cite{LeePickett-Inf-LNO:04, Botana-Inf-Nickelates:19, BernardiniCano:20}.\vspace*{-1.0em}}
	\label{fig:Bands}
\end{figure*}

The perovskite film exhibits considerable octahedral rotations
that extend into the STO substrate owing to the octahedral connectivity [Fig.~\ref{fig:Structures-ElStr}(b)].
Specifically at the interface, we obtain an apical Ni-O-Ti bond angle of $155^\circ$
and basal Ni-O-Ni and Ti-O-Ti bond angles of $155.5^\circ$ and $157.7^\circ$, respectively.
The rotational pattern of bulk NdNiO$_3$ ($a^-a^-c^+$)
is reflected in the film geometry.
In contrast, in the infinite-layer heterostructure, the small intrinsic octahedral rotations of STO are removed near the interface, and the NiO$_4$ squares show no rotations around the $c$ axis [Fig.~\ref{fig:Structures-ElStr}(a)].
% (although the ionic relaxation was initiated with considerable rotations).

For $n=2$,
the central NiO$_2$ layers are coplanar, whereas the surface (interface) layer is buckled, the Ni ion being vertically displaced outwards (inwards) from the respective oxygen layer by $\Delta z = 0.29~\AA$ ($-0.17~\AA$),
a response to the internal electric field build-up in the polar films [Fig.~\ref{fig:Structures-ElStr}(a,c)].
The distance between the surface NiO$_2$ layer and the subsurface Nd layer is considerably contracted and amounts to only $1.17~\AA$.
The Ti ions in the STO substrate show a sizeable inwards vertical ferroelectric-like displacement
(away from the interface), particularly at the interface ($\Delta z \sim -0.35~\AA$), that decays exponentially as
$\Delta z \approx -0.35~\AA \Multp \exp (- d / 7.42~\AA )$ with the distance $d$ to the interfacial TiO$_2$ layer [$S-4$; Fig.~\ref{fig:Structures-ElStr}(c)].
This trend is also expressed in strong oscillations of the apical Ti-O bond lengths around the bulk value [$1.96~\AA$; Fig.~\ref{fig:Structures-ElStr}(c)].
Our observations for $n=3$ are in sharp contrast:
In the nickelate, the Ni displacements
are smaller than $0.05~\AA$; in the STO substrate, the Ti displacements almost vanish.
This qualitatively different structural response to the polar discontinuities in infinite-layer vs.\ perovskite nickelate films points at fundamentally distinct accommodation mechanisms, which we unravel in the following.

\begin{figure}[b]
	\vspace*{-1.3em}
	\centering
	\includegraphics[]{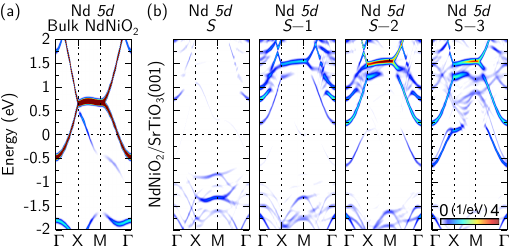}
	\caption{The self-doping Nd $5d$ states in bulk NdNiO$_2$\,(a) are shifted to higher energies and largely depleted in NdNiO$_2$/STO(001) (b).}
	\label{fig:Nd5d}
\end{figure}

% \section{Electronic reconstruction}

\textit{Electronic reconstruction. ---}
For infinite-layer nickelate films ($n=2$), we observe a considerable electronic reconstruction.
As shown in the layer-resolved density of states and charge difference relative to bulk [Fig.~\ref{fig:Structures-ElStr}(d,f)],
the polar discontinuities at the interface and the surface induce a substantial charge transfer,
expressed in a depletion at the Ni sites (i.e., of the delocalized NiO$_2$ bands), in particular at the surface, whereas the Ti ions gain charge in the localized $3d$ states, particularly at the interface, with rapid decay into the substrate.
A 2DEG emerges at the interface
% due to occupation of Ti $3d$ states
with considerable band bending in the nickelate film and the STO substrate,
as can also be seen from
% the layer-resolved density of states [Fig.~\ref{fig:Structures-ElStr}(d)] and
the band structure [Fig.~\ref{fig:Bands}(a)].
The strong Ti $3d$ occupation at the interface is in stark contrast with (LAO)$_4$/STO(001) [Fig.~\ref{fig:Structures-ElStr}(f)], which is just at the verge of a metal-insulator transition~\cite{PentchevaPickett:09},
and surprising in view of the metallic screening of NdNiO$_2$~\cite{Li-Supercond-Inf-NNO-STO:19}.
Already for ultrathin (NdNiO$_2$)$_1$/STO(001) films, we observe a 2DEG, albeit with reduced Ti $3d$ occupation ($-0.076~e^-$ at the interface).
In contrast, increasing the film thickness from 4 to 6 layers induces only negligible changes in the surface and interface electronic structure.
Since the apical Ti-O bond length at the interface is contracted [$1.86~\AA$, Fig.~\ref{fig:Structures-ElStr}(c)] relative to STO bulk ($1.96~\AA$), predominantly the $3d_{xy}$ orbital gets occupied due to its lowered energy ($\sim -0.55$~eV at the $\Gamma$ point).
This can clearly be seen in the distribution of electron density [Fig.~\ref{fig:Structures-ElStr}(d)] and resembles the situation in LAO/STO(001)~\cite{PentchevaPickett:08, PentchevaPRL:10}.
This orbital order persists for three layers and then develops into a uniform occupation of the $t_{2g}$ manifold.
In contrast, we find no tendency towards 2DEG formation in the perovskite ($n=3$) heterostructures and only negligible charge redistribution, as shown for the representative (NdNiO$_3$)$_4$/STO(001) system in Fig.~\ref{fig:Structures-ElStr}(e,g), despite a clearly visible electric field in the layer-resolved density of states of the film.

Further, we observe a high Ni $e_g$ orbital polarization throughout the infinite-layer nickelate films [Fig.~\ref{fig:Structures-ElStr}(f)] defined from the orbital occupations~$n$ by
%
%\begin{equation*}
%P = \frac{ n(3d_{x^2-y^2}) - n(3d_{z^2}) }{ n(3d_{x^2-y^2}) + n(3d_{z^2}) }
%\end{equation*}
%
$P = ( n_{3d_{z^2}} - n_{3d_{x^2-y^2}} ) / ( n_{3d_{z^2}} + n_{3d_{x^2-y^2}} )$
that increases from the interface ($P\!\sim\!12\%$) to the surface ($P\!\sim\!35\%$).
Since the $3d_{z^2}$ orbital is fully occupied,
these values reflect the approximately half-filled $3d_{x^2-y^2}$ orbital ($3d^9$ configuration).
They exceed attainable values in LaNiO$_3$/LaAlO$_3$(001) superlattices~\cite{ABR:11, WuBenckiser:13, GeislerPentcheva-LNOLAO-Resonances:19, Viewpoint:19},
where a low orbital polarization has been discussed as a major hindrance for superconductivity.
For reference, bulk NdNiO$_2$ shows an orbital polarization of $17\%$.
Interestingly, even the interfacial Ti ions exhibit a finite orbital polarization ($P\!\sim\!12\%$) due to the partial occupation of a hybridized Ni~$3d_{z^2}$--Ti~$3d_{z^2}$ interface state [$-0.65$~eV at the $\Gamma$~point, Fig.~\ref{fig:Bands}(b)].
Again, the behavior of perovskite films is distinct, with only moderate Ni orbital polarization ($P\!\sim\!11\%$) exclusively at the surface [Fig.~\ref{fig:Structures-ElStr}(g)] and otherwise largely degenerate $e_g$ states.

While the variation of $P$ directly relates to the electrostatic doping,
its high value is linked to a depletion of the self-doping Nd $5d$ states [Fig.~\ref{fig:Nd5d}(a)].
The layer-resolved density of states [Fig.~\ref{fig:Structures-ElStr}(d)] and band structure [Fig.~\ref{fig:Nd5d}(b)] for $n=2$ show that the Nd $5d$ states are largely shifted to higher energies, e.g., $\sim\!2$~eV above $E_\text{F}$ near the surface.
Merely slight contributions to the hybridized interface state can be observed.
Therefore, their involvement in the conductivity of a real heterostructure is much smaller than expected from bulk,
which considerably enhances the similarity of infinite-layer nickelates to cuprates.
Hence, the NdNiO$_2$/STO(001) Fermi surface is substantially reconstructed with respect to bulk [Fig.~\ref{fig:Bands}(b)],
being composed predominantly of Ni $3d_{x^2-y^2}$ and $\Gamma$-centered Ti $3d$ (2DEG) sheets.

\begin{figure}
	\centering
	\includegraphics[]{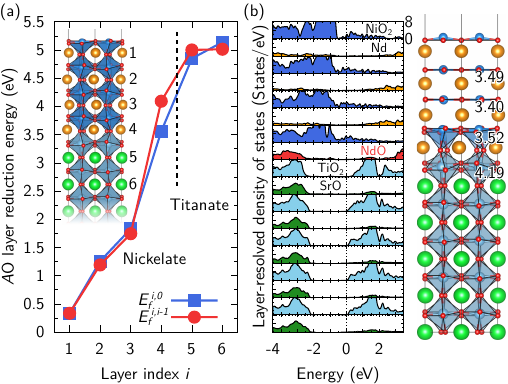}
	\caption{(a)~Energies $E_f^{i,0}$ and $E_f^{i,i-1}$ (per formula unit; see text) required to reduce different $A$O layers in (NdNiO$_3$)$_4$/STO(001) heterostructures ($A=$~Nd, Sr). (b)~Electronic structure and optimized geometry of (NdNiO$_2$)$_3$/(NdNiO$_3$)$_1$/STO(001). The oxidized layer at the interface (marked in red) inhibits the formation of a 2DEG. The Nd $5d$ self-doping is simultaneously suppressed.\vspace*{-1.0em}}
	\label{fig:OxEner}
\end{figure}

\textit{Oxygen deintercalation. ---}
We now address the topotactic oxygen deintercalation reaction from the perovskite to the infinite-layer phase.
% In order to shed some light on...
As a first approach,
Fig.~\ref{fig:OxEner}(a) displays single-layer reduction energies (i.e., apical oxygen vacancy layer formation energies), defined from DFT total energies as
%
%\begin{equation*}
$E_f^{i, 0} = E_{\text{single layer}~i~\text{reduced}}^{} - E_{\text{ideal perovskite film}}^{} + \mu_\text{O}^{}$
%\end{equation*}
%
with $\mu_\text{O} = \frac{1}{2} E_\text{O$_2$}$, i.e., in the oxygen-rich limit.
The formation energies are lowest at the surface ($\sim 0.5$~eV)
and increase to $\sim 1.8$~eV in the third layer.
At the interface, the oxygen ions are moderately bound ($\sim 3.5$~eV),
and very strongly in the STO substrate ($> 4.8$~eV)
despite the possible accommodation of electrons in the nickelate.
This is in line with an oxygen vacancy formation energy of $\sim 5.5$~eV in bulk STO~\cite{CurnanOxVac:14}
and comparable to values near the LAO/STO(001) interface~\cite{YuZunger-OxVac-LAOSTO:14}.
The values in nickelates are generally lower (e.g., bulk LaNiO$_3$: $2.8 \pm 0.2$~eV~\cite{LNO-OxVac-Beigi:15}, (LaNiO$_3$)$_3$/(LAO)$_1$(001) superlattices on STO(001): $2.3$~eV~\cite{GeislerPentcheva-LNOLAO-Resonances:19}).
The nickelate layers are therefore easily reduced,
whereas oxygen deintercalation in the STO substrate is inhibited by the high formation energies.

In a second and more realistic approach, we model successive layer-by-layer reduction,
%
% \begin{equation*}
$E_f^{i, i-1} = E_{\text{topmost}~i~\text{layers reduced}}^{} - E_{\text{topmost}~i-1~\text{layers reduced}}^{} + \mu_\text{O}^{}$,
% \end{equation*}
%
which largely concurs with the results of the first approach.
Peculiarly, we find that the oxygen binding is enhanced at the interface if the above nickelate layers are already reduced.
This leads to the surprising situation that the interfacial Nd layer may retain its oxygen under appropriately chosen experimental conditions.
Figure~\ref{fig:OxEner}(b) shows the corresponding electronic structure and optimized geometry.
The single non-reduced (i.e., perovskite) layer at the interface inhibits the formation of the 2DEG.
Instead, the STO conduction band aligns with the Fermi energy, and only little band bending can be observed directly at the interface.
No layer exhibits Nd $5d$ states near the Fermi energy,
realizing formally the $3d^9$ configuration of cuprates.
While for ideal infinite-layer films the octahedral rotations are removed at the interface [Fig.~\ref{fig:Structures-ElStr}(a)], here they are enhanced even beyond values of perovskite films.
The Nd layer distances show ionic relaxations of comparable magnitude to the ideal infinite-layer case [Fig.~\ref{fig:OxEner}(b)].
The NdO-Nd layer distance at the interface ($3.52~\AA$)
is considerably larger (smaller) than for ideal infinite-layer (perovskite) films
and thus may act as a fingerprint in transmission electron microscopy
to detect the interface layer stacking even if oxygen sites are difficult to resolve.

\textit{Summary. ---}
We investigated the electronic reconstructions in NdNiO$_n$/SrTiO$_3$(001) ($n=2,3$) driven by interface polarity from first-principles.
The results show that the polar discontinuities at the interface and the surface considerably affect the electronic structure (e.g., Ni and Ti valence, orbital polarization) throughout infinite-layer nickelate films and several layers (at least 5 unit cells) into the SrTiO$_3$ substrate.
Moreover, the accommodation mechanism is fundamentally different for infinite-layer and perovskite nickelates,
with the formation of an interfacial Ti $3d$ two-dimensional electron gas in the former case.
This implies that modeling the epitaxial films simply as strained bulk is only of limited relevance.
Hence, interface polarity emerges as a key aspect in understanding superconductivity in infinite-layer nickelates and requires further consideration in future studies.
Specifically,
while the depletion of the Nd $5d$ states in the nickelate films clearly enhances the similarity to cuprates,
we highlight the necessity to address the possible involvement of the two-dimensional electron gas in SrTiO$_3$.
Parallels to other superconducting systems such as LaAlO$_3$/SrTiO$_3$(001)~\cite{Reyren:07} and
FeSe/SrTiO$_3$(001)~\cite{Ge-FeSe-STO-SC:15} suggest more general implications for the superconductivity in polar materials on nonpolar substrates.

% \section{Acknowledgments}

\begin{acknowledgments}

This work was supported by the German Research Foundation (Deutsche Forschungsgemeinschaft, DFG) within the SFB/TRR~80 (Projektnummer 107745057), Project No.~G3.
Computing time was granted by the Center for Computational Sciences and Simulation of the University of Duisburg-Essen
(DFG Grants No.~INST 20876/209-1 FUGG and No.~INST 20876/243-1 FUGG).

\end{acknowledgments}

\vspace*{-1.5em}

% \bibliography{BibTeX/Dissertation,BibTeX/dft,BibTeX/MnSi,BibTeX/FeSi,BibTeX/GaAs,BibTeX/TM-in-Si,BibTeX/Heusler,BibTeX/ZnO,BibTeX/Oxides}


\end{document}